# Resonant scattering of spin waves from a region of inhomogeneous magnetic field in a ferromagnetic film

M.P.Kostylev [a][b] [1][2][3], A.A.Serga [a][1][4], T.Schneider[a], B.Leven[a], B.Hillebrands[a] and R.L.Stamps[b]

[a]*Technische Universität Kaiserslautern, Department of Physics and Forschungsschwerpunkt MINAS, D - 67663 Kaiserslautern, Germany*

[b]*University of Western Australia, School of Physics, M013, Stirling Highway, Crawley, WA 6009, Australia*

Abstract: The transmission of a dipole-dominated spin wave in a ferromagnetic film through a localised inhomogeneity in the form of a magnetic field produced by a dc current through a wire placed on the film surface was studied experimentally and theoretically. It was shown that the amplitude and phase of the transmitted wave can be simultaneously affected by the current induced field, a feature that will be relevant for logic based on spin wave transport.

The direction of the current creates either a barrier or well for spin wave transmission. The main observation is that the current dependence of the amplitude of the spin wave transmitted through the well inhomogeneity is non-monotonic. The dependence has a minimum and an additional maximum. A theory was constructed to clarify the nature of the maximum. It shows that the transmission of spin waves through the inhomogeneity can be considered as a scattering process and that the additional maximum is a scattering resonance.

---

[1] Both authors contributed equally to the work
[2] On leave from St.Petersburg Electrotechnical University, St.Petersburg, Russia
[3] Electronic address: kostylev@cyllene.uwa.edu.au
[4] Electronic address: serga@physik.uni-kl.de

.

## Introduction

The topic of dipole-dominated spin waves (SW) in confined geometries, such as ferromagnetic stripes and dots, is currently receiving a great deal of attention because of possible applications for data storage and processing (see, e.g., [1] and extensive literature referenced wherein). The field has grown dramatically in the past few years, due in part to advances in nanoscale engineering that make it now possible to pattern periodic arrays of elements sufficiently dense that elements interact via stray dipolar fields [2-7]. One consequence is that propagating collective spin-wave modes supported by dynamic dipole fields can be experimentally observed and studied [8]. Understanding the dynamics of these excitations is important for a number of phenomena, including fast field or current driven switching.

Propagating collective modes in these systems can be considered as a particular case of the more general phenomenon of scattering of a spin wave from a large inhomogenity in a planar geometry. Typical length scales mean that spin waves can tunnel between and through elements, or exist as confined modes within elements.

The first studies of spin wave scattering appeared in the 1980's, with an emphasis on refraction effects (see e.g. [9-11] and other numerous works by the same authors). Scatterers in these and other studies were constructed applying slowly spatially varying static magnetic field, or by depositing thin metallic layers on the surface of the film [11-18]. Scattering from other static inhomogenities in form of periodical variations like saturation magnetization or magnetostriction were also studied [19]. Some years ago, Bragg scattering from a spatially modulated magnetic field was reported [20-21].

More recently tunneling of dipole-dominated spin waves through a reststrahl region created by a locally applied magnetic field was demonstrated [22]. This region behaves much like a barrier to spin wave propagation, and can be controlled through the magnitude of the magnetic field creating the inhomogeneity. In contrast to the previous studies in this work the inhomogeneity was *highly localized* in a sense that its length was of several wavelengths or less of the spin wave incident onto it. For this purpose the local magnetic field is created by a d.c. current flowing through a wire of diameter in the micrometers placed on top of the film. In this way it is also possible to control the amplitude of the transmitted wave electrically (an example is sketched in Fig. 1a). Furthermore, the magnetic field can be modulated on nanosecond time scales. This makes the phenomena very interesting from point of view of applications, especially for signal processing at GHz frequencies.

What is most interesting perhaps is that either a barrier or a well can be created simply by changing the direction of the current. In particular, in the case of backward volume magnetostatic waves (BVMSW) a barrier is produced with a current that subtracts from the local field in the magnetic film and a well is produced by a current that produces a field which adds to the local field in the film. These effects are illustrated in Figs 1b and 1c. Propagation of BVMSW through the barrier is not possible except via tunneling [22], but propagation across the well introduces a phase shift and partial reflection (except for resonances which are discussed in detail later).

The purpose of this paper is to investigate experimentally and theoretically the transmission of dipole-dominated spin waves in a ferromagnetic film through a nonperiodic *highly localized* inhomogeneity in a ferromagnetic film. Whereas we are primarily concerned with inhomogeneities created by a dc current through a thin wire on the surface of a magnetic film, our results apply generally for any one-dimensional scalar inhomogeneity. We treat the propagation of spin waves through a region of magnetic inhomogeneity as a one-dimensional scattering problem. By analogy to quantum mechanical scattering of a particle from a potential well, scattering resonances can take place for certain spin wave wavelengths. If the well has smooth boundaries, as created by the Oersted field of the wire, the resonance condition is not trivial.

A new possible application of linear spin waves is the recently proposed SW logic [23, 24]. The logic is based on the control of the spin wave phase. In the latter paper the control achieved by varying the static magnetic field produced by a dc current through a wide magnetic stripe placed on the ferromagnetic-film sample. This allowed construction of a NOT logical gate. Both [23] and [24] used the same idea of wave interferometer for transformation of phase modulation of spin wave induced by the d.c. current into amplitude modulation of the device output signal.

A more direct way would be to directly control the spin wave amplitude. The scattering/tunneling of spin waves through a highly localized inhomogeneity provides such a possibility. Indeed, the experimental structure in Ref. [22] represents a NOT-gate. However to construct more sophisticated universal logical gates, like NAND-gate, a control of both phase and amplitude is needed simultaneously. In this regard the focus of this paper is on how current controlled tunneling and transmission affects both amplitude and phase of the scattered spin-wave.

The paper is organized as follows. In Section 1 we describe the results of experimental investigation of BVMSW propagation through an inhomogeneity of static magnetic field in an yttrium iron garnet film. We show that the dependence of amplitude of the wave transmitted through the localized inhomogeneity depends non-monotonically on the inhomogeneity height. To find the origin of this unexpected non-monotonic behavior in Section 2 we construct a theory of dipole-dominated spin wave scattering from a 1D inhomogeneity. In Section 3 the theory is applied to explain this and other experimentally observed peculiarities of BVMSW propagation through the inhomogeneity. Appendices 1 and 2 contain details of derivation of the final equations given in Section 2 and used for calculations in Section 3.

**1. Experiment**

We consider the structure shown in Fig. 1a. Microwave spin wave packets in a yttrium iron garnet film with thickness 4.9 μm are excited by microwave current pulses in a strip-line transducer. They are detected by a second transducer placed at 6 mm apart from the first one. Both a homogeneous external field $\mathbf{H}_s$ and the static magnetization $\mathbf{M}_s$ are oriented in the plane of the film parallel to the propagation direction of spin waves, $z$. The dynamic magnetization has an in-plane component, $m_y$ and an out-of-plane component, $m_x$. The propagation direction relative to the saturation magnetization ensures that backward volume magnetostatic spin waves are produced which are characterized by a negative group velocity [25].

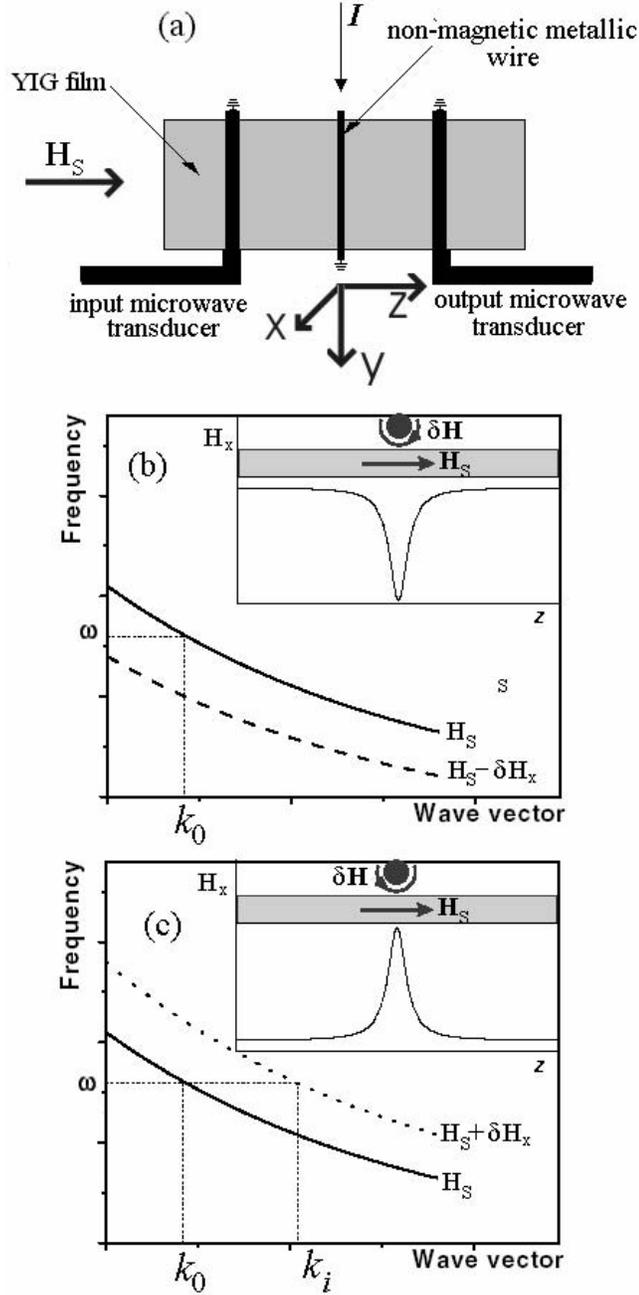

Fig. 1. Structure under investigation (a). The origin of the frame of reference coincides with the longitudinal symmetry axis of the wire.
Dip-shaped profile of the static magnetic field and BVMSW downshift of the dispersion in the regime of tunneling (b). Hump-shaped profile of the static field and upshift of the BVMSW dispersion in the regime of scattering (c). Solid lines: dispersion outside the inhomogeneity, dashed line in (b) and dotted line in (c): dispersion inside the inhomogeneity.

The microwave part of the measurement setup consists of a microwave generator and a switch, which is controlled by a pulse generator (pulse length 320-1600 ns) and connected to the input transducer. BVMSW pulses are generated with a carrier frequency $\omega/(2\pi)$=7.125 MHz, and a carrier wave vector $k_0$=49-267 rad/cm, the value of $k_0$ being determined by the dispersion relation for BVMSW in an external field.

A narrow gold wire of circular cross-section of 25 μm in diameter is mounted across the film parallel to transducers, 3 mm apart from the input transducer. The wire carries a dc current **I**. It is used to create a local inhomogeneous field $\delta\mathbf{H}(z)$. Depending on the direction of the dc current the total field is locally reduced (Fig.1b) or enhanced (Fig. 1c) by the Oersted field of the current. The negative group velocity of the backward volume magnetostatic wave [25] means that frequency decreases with increasing wavenumber. An increase of the static field shifts the manifold of allowed frequencies up (Fig. 1c), and a decrease in the field shifts the manifold down (Fig. 1b). Therefore a local decrease of the magnetic field means that the carrier frequency of the wave incident on the inhomogeneity falls outside of the spin wave manifold into the spin wave reststrahl region. Propagation in this region is not possible, but a spin wave can tunnel through this inhomogeneity if it is sufficiently narrow [22].

If the magnetic field is increased locally, the inhomogeneity takes the form of a well and merely shifts the position of the spin wave carrier frequency within the spin wave manifold. Propagation of BVMSW is allowed if the field locally does not exceed the value corresponding to the lower boundary of spin wave manifold. In order to propagate through the inhomogeneity, the carrier wave number of incident wave will adjust since the frequency remains constant, thereby resulting in a phase shift of the transmitted wave. Reflection can occur, and this decreases the power transmitted through the inhomogeneity. This provides the basic mechanism whereby both phase and amplitude can be adjusted by varying details of the current induced inhomogeneity.

To measure the transmission coefficient of the inhomogeneity, the output microwave signal from the output transducer is monitored by a microwave detector and visualized on an oscilloscope. To measure the phase of transmission coefficient as a function of the dc current through the wire $\Delta\phi(I)$ we extended the setup adding a reference circuit. This mainly consists of a directional coupler inserted between the microwave source and the microwave switch, a calibrated variable phase shifter, and a T-connector inserted between the output transducer and the microwave detector. The directional coupler couples a small portion of the incident microwave power out of the main circuit. The power passes through a calibrated variable phase shifter and a variable attenuator. The T-connector feeds the power back into the main circuit resulting in an interference of the output signal with the reference signal at the detector input. Variation of the dc current through the wire produces a change of amplitude of the interference signal. To measure $\Delta\phi(I)$ one adjusts the inserted phase via the calibrated variable phase shifter to retrieve the amplitude of the interference signal corresponding to zero dc current.

A significant dc current through such a thin wire may result in significant heating of the wire and of the film near the wire. It is well known that the temperature dependence of YIG saturation magnetization at room temperature is quite strong [26]. Account of this is necessary in order to correctly interpret experimental results. Therefore in any experiment involving increased powers precautions should be taken in order to exclude heating. In our experiments we used a procedure also used for experiments on highly nonlinear spin waves [27]. Strong signals are applied as short pulses with a small repetition rate. In this experiment to exclude heating the dc current and the microwave input signal were applied as pulses with a repetition rate of 1 ms. Intentionally the dc pulses were made shorter than the spin wave pulses. A dc pulse was applied when the central part of a spin wave pulse was beneath the wire. In such a way we were able to compare the central part of the output spin wave pulse affected by the current through the wire with the leading and trailing parts.

In this pulse regime one has two characteristic times. The first one is the transition time in the dc current circuit. The second one is the temporal length of the output spin wave pulse edges smoothed by the pulse propagation in the dispersive medium. In our experiment both times were almost equal.

We found that for very long dc pulses the restoration time to the initial levels of the amplitude and phase in the trailing part was noticeably larger than the time of transition process in the dc circuit. With the decrease of the dc pulse length the restoration time decreases and quickly reaches a stable value close to the value of the transition time in the dc circuit. This demonstrated that slow thermal processes are not relevant for these pulse lengths. These preliminary measurements allowed us to choose optimal values for lengths of the spin wave and dc pulses from the point of view of exclusion of parasitic thermal processes, and also to minimize distortion of spin wave pulses by dispersion effects. The optimal length for dc pulses was found to be 100 ns for spin wave pulses with the length 320-1600 ns long. These parameters were used in the measurements below, although for a small number of combinations of input parameters we had to increase the dc pulse length up to 150 ns to make reliable measurements.

Results are shown in Fig. 2. We found that in the tunneling regime for negative $I$-values the behavior of the amplitude of the complex transmission coefficient (shown in the upper panel of Fig. 2) is the same as that previously measured in [22]. The amplitude of the transmitted signal decreases monotonically with $|I|$. This was explained in [22] as an increase of the length of the zone prohibited for BVMSW propagation with the increase of the current magnitude. The decrease of transmission is stronger for larger incident wavenumbers.

The new result in this case is the phase characteristics. The characteristic is not linear as the phase demonstrates a tendency to saturate at large currents. Small phases are achieved by small incident wavenumbers at constant current.

Measurements with a well generated by a positive current show that scattering of BVMSW packets results in a transmission amplitude that depends non-monotonically on the magnitude of the dc current. A pronounced minimum is seen at 0.5-0.6 A. The magnitude of transmission in this minimum increases with increasing $k_0$. A weak maximum appears at a current about 1 A. For smaller wavenumbers of the incident wave ($k_0$=49-83 rad/cm) the transmission in the maximum is unity, whereas for larger wavenumbers ($k_0$=116 and 158 rad/cm) the transmission at the maximum is only partial.

The phase of the transmitted signal shown in the lower panel has a general tendency to decrease linearly with $I$, but with noticeable deviations from linearity in the vicinities of $I$-values which correspond to the minimum and the maximum (on the right hand side) in the upper figure. The depth of the minimum is inversely related to $k_0$, and corresponds to strongly nonlinear behavior of $\Delta\phi(I)$.

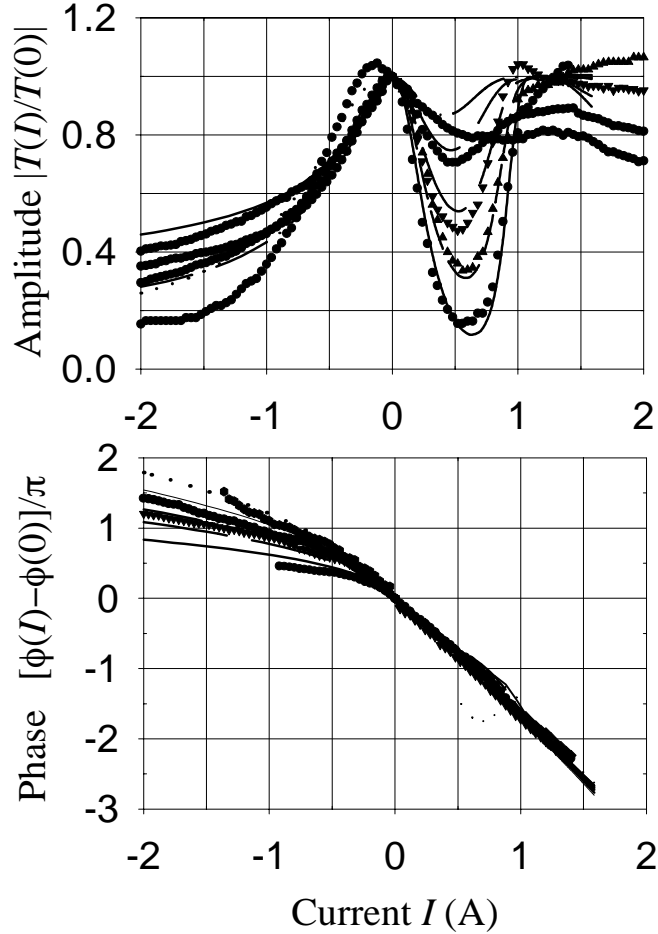

Fig. 2. Measured and calculated transmission coefficient. Upper panel shows the amplitude; the lower panel shows the phase. Symbols show experimental data. Lines show results of numerical calculation. Measurements and calculations were made for $k_0$=49, 63, 83, 116, 158, and 267 rad/cm.

Upper panel, $I>0$: the lowest curve corresponds to $k_0$=49 rad/cm, the highest one to $k_0$=158 rad/cm. $I<0$: the curves (experimental and theoretical) showing the largest transmission at $I=-2$ A are for $k_0$=49 rad/cm, those showing the lowest one are for $k_0$=267 rad/cm. The theoretical and the experimental lines for middle curves nearly coincide.

Lower panel: $I>0$: All the curves practically coincide within the graphical accuracy. $I<0$: the lowest curves (experimental and theoretical) are for $k_0$=49 rad/cm, the highest ones are for $k_0$=267 rad/cm.

As one sees from Fig. 2, the results for phase at positive currents are shown only up to $I$=1.5 A. The reason for this is that the controllable phase shifter we used in our phase measurements was able to shift the phase of the microwave reference signal only up to $2.5\pi$. For negative $I$-values this was sufficient to cover the whole range of accessible dc current values, but for the positive currents we were obliged to stop our measurements at $I=1.5$A because of the stronger slope of $\Delta\phi$ dependence on $I$. Similarly we show the amplitude characteristic for the smallest value of $k_0$ only up to 1.5 A. We were not able to measure the amplitude for larger currents because of significant oscillations in the profile of the output pulse.

Note that the measured phase characteristic allows one to independently check for the influence of ohmic heating by the dc current on the experimental results. As stated above, a local heating of the YIG film would locally reduce the film saturation magnetization, similarly to the example shown in Fig. 1b for the negative Oersted field. A decrease of saturation magnetization should shift the dispersion curve downwards, like the negative Oersted field does in Fig. 1b, opposite to the shift induced by the actual positive Oersted field of the wire (shown in Fig. 1c). The Oersted field is linearly proportional to the current, whereas the heat is proportional to $I^2$. Thus a possibility exists that for larger positive $I$-values these local shifts of the dispersion curve compensate each other, which might result in full transmission at the point of full compensation. Therefore one can suppose that the non-monotonic behavior of the transmitted amplitude showing full transmission for $I=1$ A in Fig. 2 has the simple explanation that the maximum at 1A is the point of the compensation. In such a case, the phase of the transmitted signal at the point of compensation would be close or equal to the value at $I=0$. Furthermore, the influence of the Oersted field should dominate the thermal effects at smaller currents. In the range 0.5-1 A, the entire phase characteristic would be strongly nonmonotonic with a minimum situated at the same point as for the amplitude (Fig. 2, upper panel). Since the measured phase characteristics are instead linear, this explanation cannot account for the observed nonmonotonic behavior of the amplitude.

## 2. Theory

We now construct a theory able to explain the observed amplitudes. The theory is also able to explain the measured nonlinear phase dependence on *I* in the tunneling regime ($I<0$).

The propagation of dipolar spin waves in a film geometry is complicated because of off-diagonal terms in the permeability associated with the gyromagnetic response. However for the special case of a plane wave traveling in the *z* direction (BVMSW) and incident normally on a discontinuity with translational invariance in the *xy* plane, the off-diagonal terms are not important and the wave satisfies the integral equation

$$\chi(z,\omega)^{-1} m_x(z) - \int_{-\infty}^{+\infty} 4\pi G_{xx}(z,z') m_x(z') dz' = A\delta(z-z_0). \tag{1}$$

Here $\chi(z,\omega)$ is the diagonal term of the microwave magnetic susceptibility tensor $\hat{\chi}$ [28,29], and *z* is the direction of BVMSW propagation. The kernel $4\pi G_{xx}(z,z')$ is the out-of-plane diagonal component of an approximate quasi-1D Green's function $\mathbf{G}(z,z')$ for the magnetostatic field produced by sources in a planar geometry. The Green's function was first obtained in [30] and the diagonal out-of-plane component has the form

$$4\pi G_{xx}(z,z') = \frac{2}{L} \ln \frac{(z-z')^2}{L^2 + (z-z')^2}, \tag{2}$$

where $L$ is the film thickness. The Dirac delta function in Eq. (1) describes the linear excitation of an incident monochromatic spin wave by a source at a point $z_0$ located far from the inhomogeneity ($z_0 \to -\infty$). $A$ is the excitation amplitude of the source. A derivation of Eq. (1) is given in Appendix A.

Propagating dipole-dominated BVMSW exist in the range $-1 < \chi(z,\omega)^{-1} < 0$ (see e.g. [25,29]). Outside this range only evanescent waves can exist [22]. In the experimental situation $\hat{\chi}(\omega,z)$ depends on position $z$ because of the spatial variation of the field inhomogeneity. The total static magnetic field is a superposition of the applied field $\mathbf{H}_s = H_s \mathbf{e}_z$ and $\delta \mathbf{H}(z)$, the field created by the dc current in the neighboring wire. For simplicity we consider only the in-plane component of this field so that the total field is entirely along the $z$ direction with magnitude $H(z) = H_s + \delta H_z(z)$. Far from the wire the field is due only to $H_s$ so that $H(z \to \pm\infty) = H_s$.

The value of the $z$-component of the wire induced field averaged through the film thickness is:

$$\delta H_z(z) = Y(z) I , \qquad (3)$$

where $Y(z)$ is the profile of the field created by the current. One can show that for cylindrical wire,

$$Y(z) = \frac{1}{5L} \ln \frac{z^2 + (r+d+L)^2}{z^2 + (r+d)^2} , \qquad (4)$$

where *I* is the dc current, *r* is the wire radius, and *d* is the nearest distance between the surfaces of the wire and film.

## Numerical solution of the equations of motion

First we solve Eq. (1) numerically. Because in the experiment spin waves are not monochromatic, we first generalize Eq. (1) as an inhomogeneous time dependent integro-differential equation. This allows us to make calculations for pulsed spin wave propagation.

We make calculations for incident spin wave pulses 100-300 ns of duration and register the amplitudes of transmitted signals in the center of the pulses. Our first main observation is the amplitudes of the transmitted and reflected pulses of such length far away from the pulse edges do not depend on the pulse duration, thus in the later analytical treatment for the sake of simplicity we may consider monochromatic spin waves.

The solid lines in Fig. 2 show the results of the numerical solution. As in the experiment amplitudes and phase from the center of pulses are shown. As seen from the figure, there is a good agreement between the simulation and experiment. A free parameter used for the fit was spacing, taken as *d*=10 μm (4) between the wire and the film surface.

In order to achieve good agreement, the actual values of positive currents *I* were reduced by 0.881. Reasons for the reduction include neglect of the out-of-plane component of the current created field $\delta H_x(z)$ and use of a one-dimensional Green's function of the dipole field $G_{xx'}(z,z')$. These two factors should increase the reflection from the inhomogeneity, and thereby enhance the effect of *I* on transmission. The reduction may also describe the influence of residual heating effects. As a result our numerical treatment probably underestimates the current.

Despite underestimating the current, the model of Eq. (1) is able to describe the main effects observed in the experiment. Calculation of power carried by pulses shows that the sum of powers carried by the transmitted and reflected pulses is equal to the difference of the power of the input spin wave pulse and the power lost due to magnetic damping. Energy is conserved and we can conclude that the minimum of propagation in the calculated dependences corresponds to the maximum of reflection.

Finally we note that the numerical solution of the time independent Eq. (1) introduces spurious full reflection from the integration boundaries. The time-dependent equation for pulses we used allows us to separate the transmitted and the reflected pulses through time delays. We can therefore identify unambiguously transmitted and reflected power without significant losses to spurious reflections.

## Integral equation formulation of the scattering problem

Additional insight the problem is obtained using an alternative solution method.

We consider monochromatic spin waves and assume that $\delta H_z(z) \ll H_s$. This allows us to transform (1) into

$$m(z) = I \int_{-w/2}^{w/2} G_{exc}(z,z') \delta \upsilon(z',\omega) m(z') \mathrm{d}z' + \exp\left(i k_0^c z\right). \tag{5}$$

The derivation of Eq. (5) is shown in Appendix B. In this equation $I\delta\upsilon(z,\omega) = [\chi(z,\omega)^{-1} - \chi(z=\pm\infty,\omega)^{-1}]/(4\pi) - \left(O(I/Hs)^2\right)$, $k_0^c$ is the complex wavenumber of the incident spin wave, and $-w/2 < z < w/2$ is the region of localization of the current created field. An expression for $G_{exc}$ is given in Eq. (48) and determined from Eq. (39). A key point for our analysis is that the finite range of integration in (5) results in a discrete spectrum of eigenmodes.

Equation (5) for the spin wave amplitude is now analogous to the Green's function formulation of the direct scattering problem in quantum mechanics (see e.g. [31]), and $I\delta\upsilon(z,\omega)$ plays the role of a scattering potential.

### Green's function and Born approximation

Equation (5) represents a sum of the incident and scattered fields of the form $m(z) = S(z) + \exp(ik_0^c z)$, where the scattered field $S(z)$ is the integral $S(z) = I \int_{-w/2}^{w/2} G_{exc}(z,z')\delta\upsilon(z',\omega)m(z')dz'$. Far away from the inhomogeneity the scattered field can be decomposed into a sum of two waves: $S(z) = S_+(z) + r(z)$ where $S_+(z)$ is the forward scattered wave and $r(z)$ is the back scattered (reflected) wave.

Later we will solve Eq. (5) exactly, but some insight can be gained by using Born's approximation [31]. In the first Born's approximation to obtain amplitudes of the transmitted, reflected and scattered waves it is necessary to place the observation points far away from the inhomogeneity. This reduces the Green's function to a simple expression (50). The scattered field approximated to first order is found by calculating $S(z)$ using $m(z) = \exp(ik_0 z)$. Neglecting spin wave losses by setting $\upsilon_0''=0$ (see Appendix B for the definition of the loss parameter $\upsilon_0''$), the transmitted and reflected amplitudes in the first Born approximation far from the scattering inhomogeneity are

$$S_+(z) = i\frac{2I}{L}\exp(i|k_0|z)\int_{-w/2}^{w/2}\delta\upsilon(z',\omega)dz', \quad z \gg w/2,$$
$$r(z) = i\frac{2I}{L}\exp(-i|k_0|z)\int_{-w/2}^{w/2}\delta\upsilon(z',\omega)\exp(2i|k_0|z)dz', \quad z \ll w/2. \tag{6}$$

The amplitude of the transmitted wave $S_+(z)$ is linearly proportional to the area of the inhomogeneity profile $\Xi = \int_{-w/2}^{w/2}\delta\upsilon(z',\omega)dz'$. This quantity is the zeroth order spatial Fourier component of the inhomogeneity profile. The reflection amplitude is proportional to the $2k_0$ Fourier component of the inhomogeneity profile:

$Q = \int_{-w/2}^{w/2}\delta\upsilon(z',\omega)\exp(-2i|k_0|z)dz'$; i.e., the first resonant Bragg backscattered wave.

Both integrals for the inhomogeneity profile can be calculated. Using the notation in Appendix B, the results are:

$$\Xi = \frac{2}{5}\pi\eta(\omega),$$
$$Q = \frac{2}{5}\pi\eta(\omega)\left[1-\exp(-2|k_0|L)\right]\exp[-2|k_0|(r+d)]/[2|k_0|L]. \tag{7}$$

The transmission coefficient is $|T| = |S_+(z) + \exp(ik_0 z)| = \sqrt{1+(2I\Xi/L)^2}$. For $I \neq 0$ we have the unphysical result that $|T|>1$ and the first Born approximation clearly fails even for small $I$. Nevertheless, examination of the upper panel in Fig. 3b shows that the first order Born approximation estimates $|S_+(z)|$ well for small *I*. The problem with the transmitted coefficient *T* is because of the incorrect treatment of the phase of $S_+(z)$ This is illustrated in the lower panel of Fig. 3b.

For $|k_0|L \ll 1$ the quantity $\left[1 - \exp(-2|k_0|L)\right]/(2|k_0|L)$ appearing in *r(z)* reduces to $1 - 2|k_0|L$. Hence *Q* is an increasing function of *I* and a decreasing function of $k_0$ and approaches zero as $k_0 \to \infty$. If there is no dissipation we require $|T| = \sqrt{1^2 - |R|^2}$. Hence the transmission coefficient decreases with *|I|* and increases with increasing $k_o$. This behavior is in qualitative agreement with experiment and also with results from the more rigorous solution of (5) as depicted in Fig. 3b. Note that the range of validity is $|I| \leq L/(2Q)$, where the reflection coefficient remains less than 1. Furthermore, the experiment is able to probe the range $k_0=0$ to $k_0=200$ rad/cm which means that our Born approximation is valid only for currents less than 0.38 A at most. This is much smaller range than accessible in existing experiments.

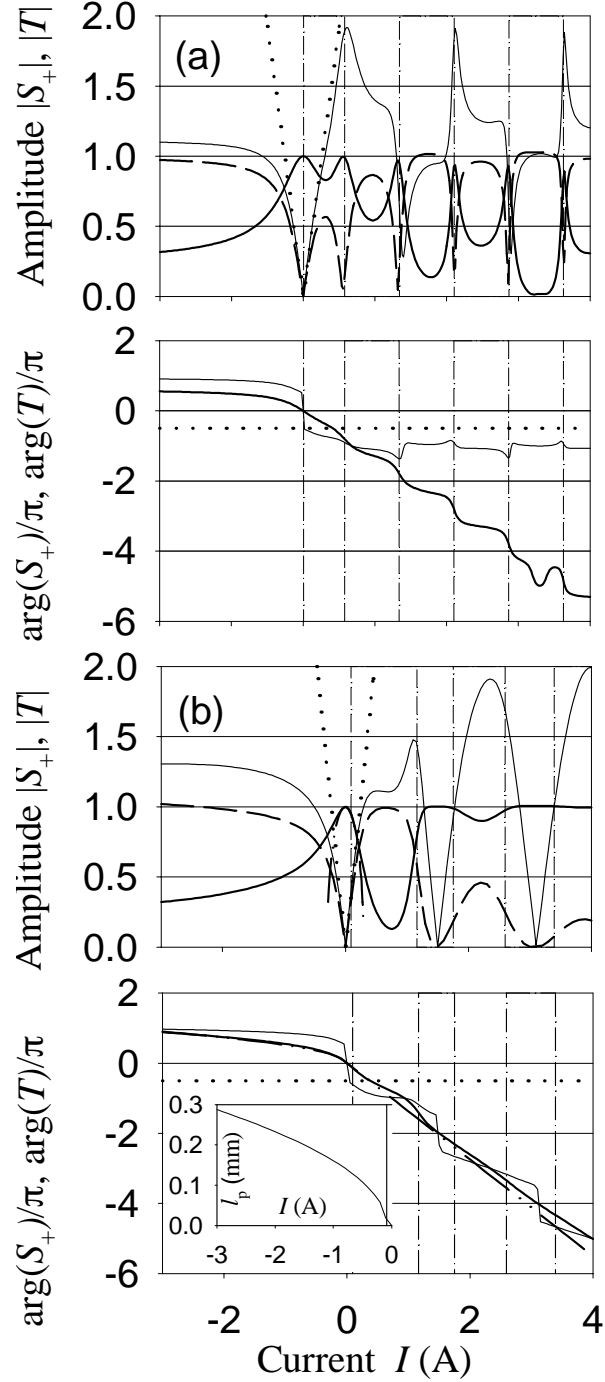

Fig. 3. Amplitudes (upper panels) and phases (lower panels) of transmitted (solid line), reflected (dashed line) and scattered (thin solid line) waves for a rectangular-shaped inhomogeneity (a) and for the wire with current (b). For comparison the dotted lines show the amplitude and phase of the forward scattered wave calculated in the first Born's approximation, and the dash-dot-dotted line in the lower panel of Figure (b) is the WKB approximation for the phase of transmission coefficient. The dash-dot-dotted line in the upper panel of Figure (b) is the transmission coefficient calculated as $|T| = \sqrt{1^2 - |R|^2}$ with $R$ from the first Born's approximation. Thin vertical dash-dotted lines show the positions of the transmission resonances, as calculated from (8) for these potential shapes. Inset in the lower panel of Figure (b) shows the dependence of the length of prohibited zone on the wire current.

Exact solution to the integral equation

The difficulty with a perturbative approach is its inadequacy of describing the near field. From Fig. 2 we see that a good approximation for *T* will require at least a third order dependence on *I*, and therefore will require several terms beyond the leading one in a perturbation expansion. This is cumbersome, as one needs to use the whole Green's function for substitution into higher-order term integrals. We use instead a different method based on an eigenfunction expansion of the integral operator kernel in Eq. (5).

We first solve

$$\lambda u(z) = \int_{-w/2}^{w/2} G_{exc}(z,z') \delta \upsilon(z',\omega) u(z') \mathrm{d}z', \tag{8}$$

and the transposed operator

$$\lambda \phi(z) = \delta \upsilon(z,\omega) \int_{-w/2}^{w/2} G_{exc}(z',z) \phi(z') \mathrm{d}z' \tag{9}$$

for eigenfunctions and eigenvalues. Note that with the substitutions $u(z) = \tilde{u}(z)/\sqrt{\delta \upsilon(z,\omega)}$, $\phi(z) = \sqrt{\delta \upsilon(z,\omega)} \tilde{\phi}(z)$ both equations are seen to have the symmetric kernel $\sqrt{\delta \upsilon(z,\omega)} G_{exc}(z,z') \sqrt{\delta \upsilon(z',\omega)}$.

Once the eigenvalues and the eigenfunctions are found, the solution of the inhomogeneous equation is expressed as follows:

$$m(z) = m_{exc}(z) , \qquad (10)$$

where

$$m_{exc}(z) = \sum_{n=0}^{\infty} \frac{\int_{-w/2}^{w/2} \exp(ik_0^c z')\phi_n(z')dz'}{1 - I\lambda_n} u_n(z) . \qquad (11)$$

To obtain Eq. (11) we use the bi-orthogonality of the sets of eigenfunctions

$$\int_{-w/2}^{w/2} \phi_n(z)u_{n'}(z)dz = N_n^2 \delta_{nn'} , \qquad (12)$$

and normalize the functions such as $N_n = 1$. The solution (10) is valid inside the interval of the bi-orthogonality of the functions $u(z)$ and $\phi(z)$: $-w/2 < z < w/2$ (12). As seen from (11), $m(z)$ may depend on $I$ in a resonant way. The resonant condition is

$$\text{Re}(1/\lambda_n) - I = 0 . \qquad (13)$$

The scattered field is then the difference between the full solution Eq. (10) and the unscattered wave:

$$S(z) = m_{exc}(z) - \exp(ik_0^c z), \quad -w/2 < z < w/2 . \qquad (14)$$

In the vicinity of the incidence boundary of the inhomogeneity near $z = -w/2$, the scattered field represents only a reflected wave. Hence the solution for the reflected wave is

$$r(z) = m_{exc}(z) - \exp(ik_0^c z), \quad -w/2 < z << 0. \tag{15}$$

Similarly near $z = w/2$ only the unscattered and the forward-scattered waves are present. Here the solution for the transmitted wave is

$$t(z) = S(z) + \exp(ik_0^c z) = m_{exc}(z), \quad 0 << z < w/2. \tag{16}$$

As stated above, the solution Eqs. (14)-(16) are valid only inside the inhomogeneity. An expression valid at any $z$ is obtained by substituting Eq. (10) into Eqs. (5). The scattered field is then

$$S(z) = I \int_{-w/2}^{w/2} G_{exc}(z,z') \delta\upsilon(z',\omega) m_{exc}(z') \mathrm{d}z', \quad -\infty < z < \infty. \tag{17}$$

Furthermore,

$$r(z) = S(z), \quad z < -w/2, \tag{18}$$

and

$$t(z) = S(z) + \exp\left(ik_0^c z\right), \quad z > w/2, \tag{19}$$

with $S(z)$ from (17).

The transmission coefficients $R$ and $T$ are found by the asymptotic limit $z \ll w/2$ of Eq. (18) and $z \gg w/2$ of Eq. (19). In what follows we find the eigenfunctions $u(z)$ and $\phi(z)$ numerically on the finite interval $-w/2 < z < w/2$.

We note that in the limiting case $w \to \infty$ (which corresponds to a smooth potential like Eq. (4)) the bi-orthogonality interval is the whole $z$-axis and one should use Eq. (15) and Eq. (16), rather than Eq.(19) and Eq. (18) to calculate $T$ and $R$. In this limit $|R| = |r(-w/2)|$ and $|T| = |t(w/2)|$.

In Appendix C we derive an explicit formula having a validity range larger than that obtained in Born's approximation and free of necessity to numerically solve the eigenvalue problem (8)-(9).

## 3. Discussion

In the numerical implementation of the eigenfunction method described above, care must be taken with the finite width of the inhomogeneous region. At the boundaries of the inhomogeneity $\delta \upsilon(z,\omega)$ discontinuously changes to zero. A non-physical reflection from the boundary will appear but can be minimized by decreasing the magnitude of the jump. This is accomplished by choosing a large $w$ in Eq. (5). Since the current induced field of the thin wire is highly localized, it is not difficult to satisfy this condition in numerical calculations.

The solid line in the upper panel of Fig. 4 is the result from a calculation of the eigenvalues of Eqs. (8) and (9). For this calculation $w/2$ was set equal to $80r$. This corresponds to a jump in the inhomogeneity field $\delta \mathbf{H}(w/2) = 6 \times 10^{-4} \, \delta \mathbf{H}(0)$. Such a small jump of the field does not produce any noticeable reflections in the simulation. In the experiment the strength of dc applied current did not exceed $\pm 2$ A, so by the resonant condition Eq. (13), the relevant inverse eigenvalues are small: $-5A < \text{Re}(1/\lambda_n) < 5A$ in this particular case.

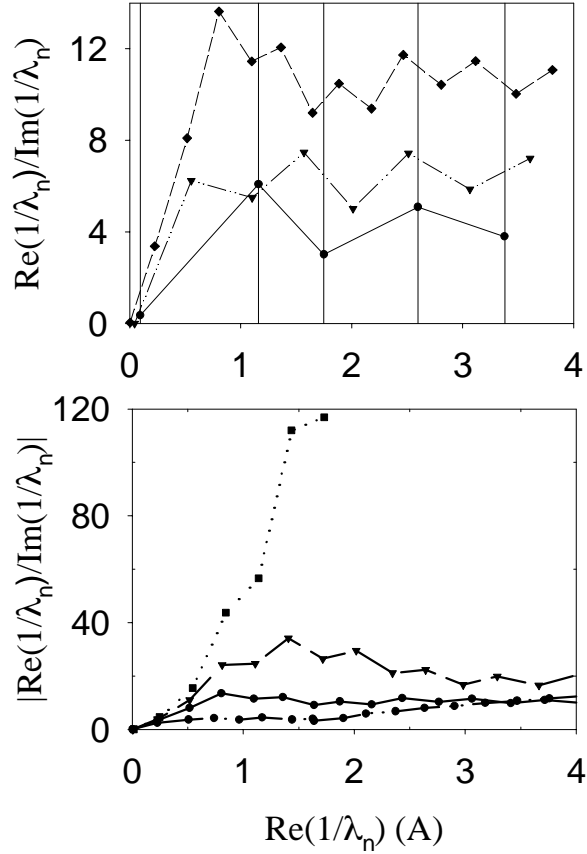

Fig. 4. Complex eigenvalues $\lambda_n$ of (8) and (9) shown as "quality factors" $\text{Re}(1/\lambda_n)/\text{Im}(1/\lambda_n)$ vs. real eigenvalue parts $\text{Re}(1/\lambda_n)$. Upper panel: solid line is for the experimental profile (4) (in terms of (21) $t=50$ μm, $l_{\text{eff}}=25$ μm). The dashed and the dash-dotted lines are for the profile (21) of lengths $t=200$ and 100 micrometers respectively. Lower panel: the same for the profile (21) with different steepness of the edges. The solid line is for the experimental edge steepness $l_{\text{eff}}=25$ μm, the dashed and the dash-dotted lines are for the edges with the effective lengths 15 and 50 micrometers. The dotted line is for the square-shaped inhomogeneity (20). The whole length of the inhomogeneity $t$ in the lower panel is 200 μm.

The eigenvalues belonging to this range are situated in the complex plane close to the real axis. As a rule the real parts are larger than the imaginary parts. As seen from Fig. 4, the inverse eigenvalues are not distributed evenly along the real axis and the distance between neighbouring points on the real axis changes non-monotonically with frequency.

In Fig. 5 the calculated eigenfunctions of Eqs. (8) and (9) are shown. The eigenfunctions of the transposed operator Eq. (9) determine the scattering efficiency since they determine the overlap integral with the incident wave in the numerator of Eq. (11). These modes are localized at the inhomogeneity and the modulus of the lowest frequency mode is very close to the profile of the inhomogeneity field given by Eq. (4). The eigenfunctions of Eq. (8) determine the amplitude of the scattered waves at $z = \pm\infty$, and hence represent the reflection and transmissions coefficients. These functions have the asymptotic form of monochromatic travelling waves at $z = \pm\infty$.

The transmission coefficient calculated from Eqs. (11) and (16) with the eigenfunctions shown in Fig. 5 is presented in Fig. 6. Experimental data and the results of numerical simulation from Fig. 2 are shown for comparison. There is a good agreement with the experiment for both calculations. The small discrepancy between the numerical simulation and the eigenfunction expansion results is due to keeping only the first two terms in $k_0$ in the expansion of $W(k)$ (47) and only the first two terms in $I$ in the expansion of the inhomogeneity profile (5).

The profiles of the amplitude and phase of the transmitted wave given by Eq. (16) is shown in Fig. 3b. The reflected wave from Eq. (15) and the forward scattered wave from Eq. (14) are also shown. The calculation is made over a large range of $I$ values in order to assess the validity of the approximations made using Eqs. (33) and (47).

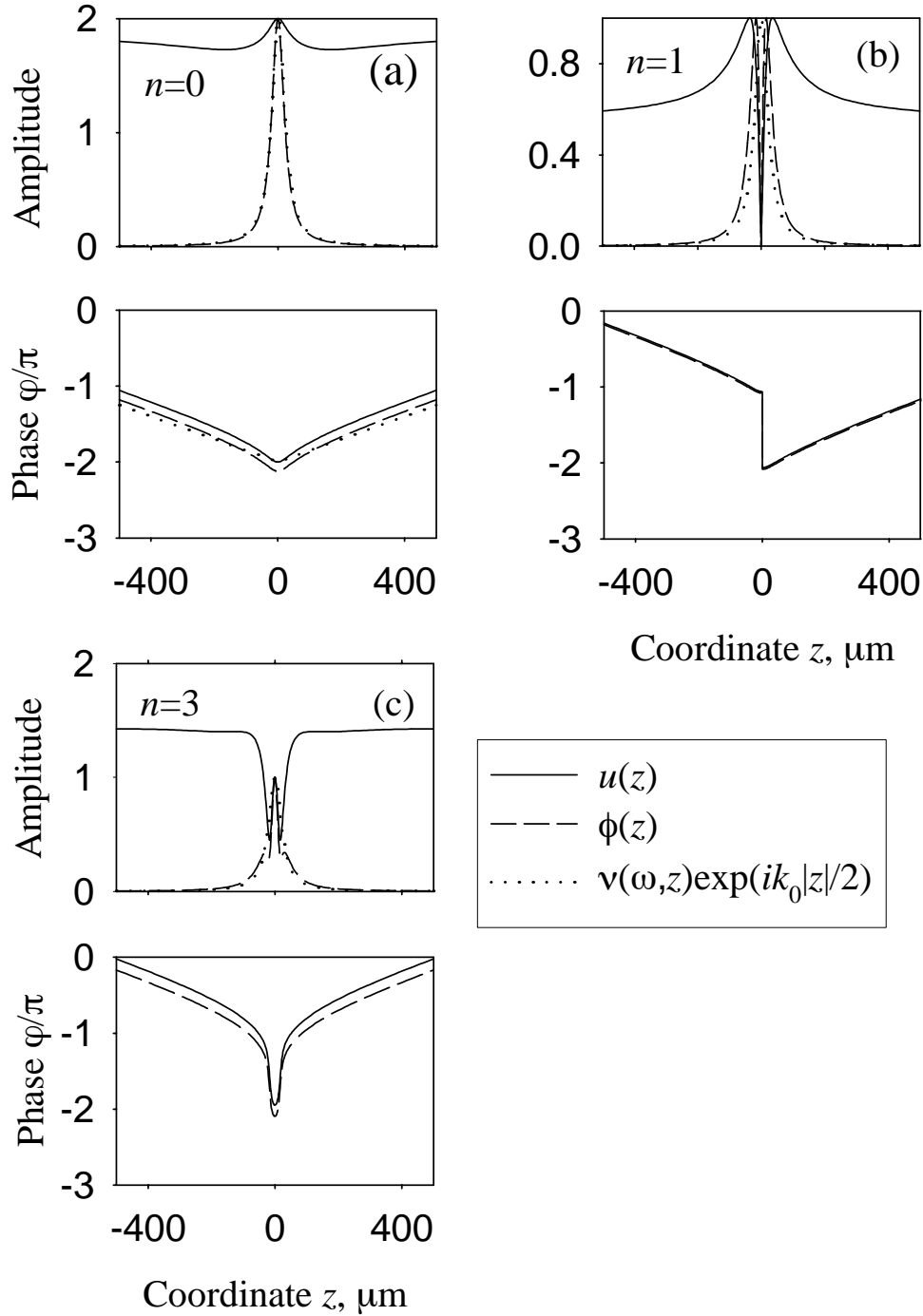

Fig. 5. Lowest eigenfunctions of (8) and (9) for $k_0=47$ rad/cm. (a), $n=0$ ($1/\lambda_0=0.098-0.279i$); (b) $n=1$ ($1/\lambda_1=1.193-0.1198i$); (c) $n=2$ ($1/\lambda_2=1.172-0.597i$). Upper panels are amplitudes of the complex eigenfunctions, and the lower panels are their phases. Solid lines show eigenfunctions of (8), and the dashed lines – those of the transposed problem (9). The dotted lines show the profile of the Oersted field (4) and the linear phase profile $\exp(i\,k_0|z|)$.

One sees in Fig. 3b that minima observed in the experimentally accessible range of small $I$ values are not unique. Weaker minima appear for larger values of $I$ as well. Vertical lines in the figure show the positions of roots of Eq. (13). The roots do not fully coincide with the positions of transmission maxima. Except the trivial root $n = 0$ at $I \approx 0$, the next roots are situated at the edges of wide plateaus of full transmission.

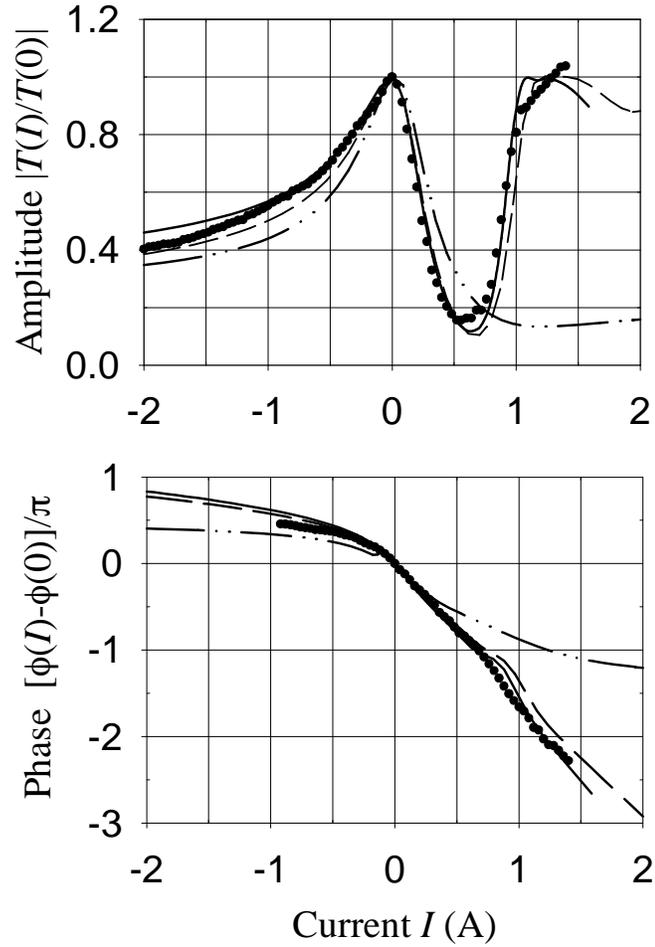

Fig. 6. Transmission coefficient. Upper panel: amplitude; lower panel: phase. Solid line: numerical calculation, dashed line: approximate analytical calculation, dash-dot-dotted line: approximated calculation by using (51); symbols: experiment. Initial wavenumber $k_0$=47 rad/cm.

## Transmission resonances for a rectangular inhomogeneity

We can identify the origin of the plateaus by solving Eq. (5) for the case of the rectangular-shaped inhomogeneity modeled by the expression below:

$$\delta v(z,\omega) = \begin{cases} \delta v(0,\omega), & -w/2 < z < w/2 \\ 0, & z < -w/2, \, z > w/2 \end{cases}, \qquad (20)$$

where $\delta v(0,\omega)$ is $\delta v(z,\omega)$ from (35) with (4) calculated by setting $z = 0$. With this profile the integral operator in Eqs. (5) with (20) reduces to an integration of $G_{exc}(z,z')$ over $-w/2 < z' < w/2$. In order to make the phenomenon more pronounced we set the length of the rectangular inhomogeneity greater than the actual wire diameter and use a large incident wavenumber: $w = 100 \, \mu$m with $k_0 = 120$ rad/cm. The inhomogeneity profile of Eq. (20) describes the magnetic field created by a dc current through a thin stripe conductor placed directly on the film surface.

The panels in Fig. 3a contain the solution of Eq. (5) for this rectangular-shaped inhomogeneity. As seen from the upper panel, scattering of BVMSW results in a set of distinct maxima in $|T(I)|$ and $|S_+(I)|$ for $I > 0$. Positions of the maxima coincide with the roots of Eq. (13). It indicates that these are indeed resonances. Since (11) is the scattering problem, we may infer that these are "transmission resonances".

Transmission resonances are created by multiple reflections from the inhomogeneity boundaries and occur when an incident wave excites unstable bound states. A transmission resonance occurs when the wave reflected from the boundaries inside the inhomogeneity interferes constructively with the incoming wave. Destructive interface leads instead to enhanced reflection.

There are similarities between the above transmission resonances and the quantum mechanical problem of scattering from a one dimensional potential well. The difference is that in the present case we cannot write the scattered spin waves as simple plane harmonic waves because of the nonlocality of the dipole interaction. The dipole interaction couples the sources inside and outside the inhomogeneity so that purely propagating harmonic solutions with $k_0$ and $k_i$ exist only far away from the boundaries. This means that boundary scattering involves a superposition of all modes and cannot be simplified at each interface into scattering between three waves as in the one dimensional quantum well problem.

In the case of rectangular inhomogeneity it is easy to analyse the transmission in terms of partial reflection of waves from the inhomogeneity boundaries. The first minimum of transmission coincides with the fist maximum of reflection (dashed line in Fig. 3a) when the partial waves reflected from the front and the rear boundaries are in phase. The maximum transmission takes place when these partial waves cancel each other at the front boundary. Since the wave inside the inhomogeneity may be reflected several times from the boundaries this forms a transmission or reflection resonance.

To gain more insight into formation of resonances, we made additional numerical calculations for spin wave propagation through the rectangular inhomogeneity in the pulse regime using the same method as in Section 3. The length of the inhomogeneity was chosen to be as long as possible (several millimetres). The phases of spin wave pulses reflected from the front edge of the inhomogeneity for $I > 0$ (reflection from a positive step on the static-field profile) and for $I < 0$ (reflection from the negative step) were calculated. We found that the phase of the reflected pulse is shifted by $\pi$ with the respect of the incident wave if $I > 0$. The phase difference is zero if $I < 0$.

The latter case models well the reflection from the rear boundary of short positive ($I > 0$) rectangular inhomogeneity in Fig. 3. The numerical calculations show that the minima of transmission correspond to a phase accumulation of an odd multiple of $\pi/2$ by the wave propagating in the forward direction inside the inhomogeneity. The wave accumulates this phase as it travels from the front boundary to the rear boundary. The wave reflected from the rear boundary accumulates the same phase on its way to the front boundary, since the internal reflection from the rear boundary results in no phase shift, as discussed before. Consequently the phase accumulated along the whole loop is an odd number of $\pi$, and the signal passed along the whole loop meets the wave reflected back from the front boundary with the phase difference equal to an even number of $\pi$.

Similarly in the maxima of transmission the phase accumulated on the length of the inhomogeneity equals to an even number of $\pi$. Consequently the partial wave going directly through it meets in phase the wave reflected first from the rear boundary and then from the front one.

By way of an optical analogy, a region of increased magnetic field acts as a region of increased refraction index for BVMSWs. This analogy does not go far however. Because of strong near fields at the boundaries due to the long-range dipole interaction a standing wave profile with a definite value of $k_i$ exists only far away from the boundaries of the inhomogeneity. This precludes, for example, the use of transfer matrices for formulating the problem.

Transmission resonances with a smooth profile inhomogeneity

Scattering resonances are not formed at all for some kinds of smooth potentials [32] and in general the problem can be complicated. In order to gain insight into the effect of the inhomogeneity profile on scattering, we have solved Eq. (5) for a modified rectangular inhomogeneity profile:

$$\delta v(z,\omega) = \begin{cases} \delta v(z+t/2-l_{eff},\omega), & z < -t/2+l_{eff} \\ \delta v(0,\omega), & -t/2+l_{eff} \leq z \leq t/2-l_{eff} \\ \delta v(z-t/2+l_{eff},\omega), & z > t/2-l_{eff} \end{cases}, \qquad (21)$$

where $\delta v(z+t/2-l_{eff},\omega)$ and $\delta v(z-t/2+l_{eff},\omega)$ are $\delta v(z,\omega)$ from Eq. (35) calculated by setting $z = z+t/2-l_{eff}$ or $z = z-t/2+l_{eff}$ in Eq. (4) respectively.

This profile has smooth boundary slopes and a plateau of constant amplitude $\delta v(0,\omega)$ of length $t-2l_{eff}$. The length of each slope measured at midheight is equal to $l_{eff} = \sqrt{(r+d_0)(r+d_0+L)}$. The entire length of the inhomogeneity from these points is $t$. A finite spacing of $d$=10 μm between the wire and the film surface is again used, giving a length of $l_{eff}$ =25 μm for a wire diameter of 25 micrometers.

Calculations were made for $t$=200, 100 and 50 μm. Results are shown in Figure 7. $|T(I)|$ exhibits a set of maxima for positive $I$ for the longest $t$. The positions of maxima coincide with the roots of Eq. (13) as expected. The positions of each second resonance for $t$=200 μm almost coincide with the positions of resonances for $t$=100 μm. At large $I$, plateaus of perfect transmission are formed instead of the peaks. The reason is that roots of Eq. (13) are situated in this case very close to each other.

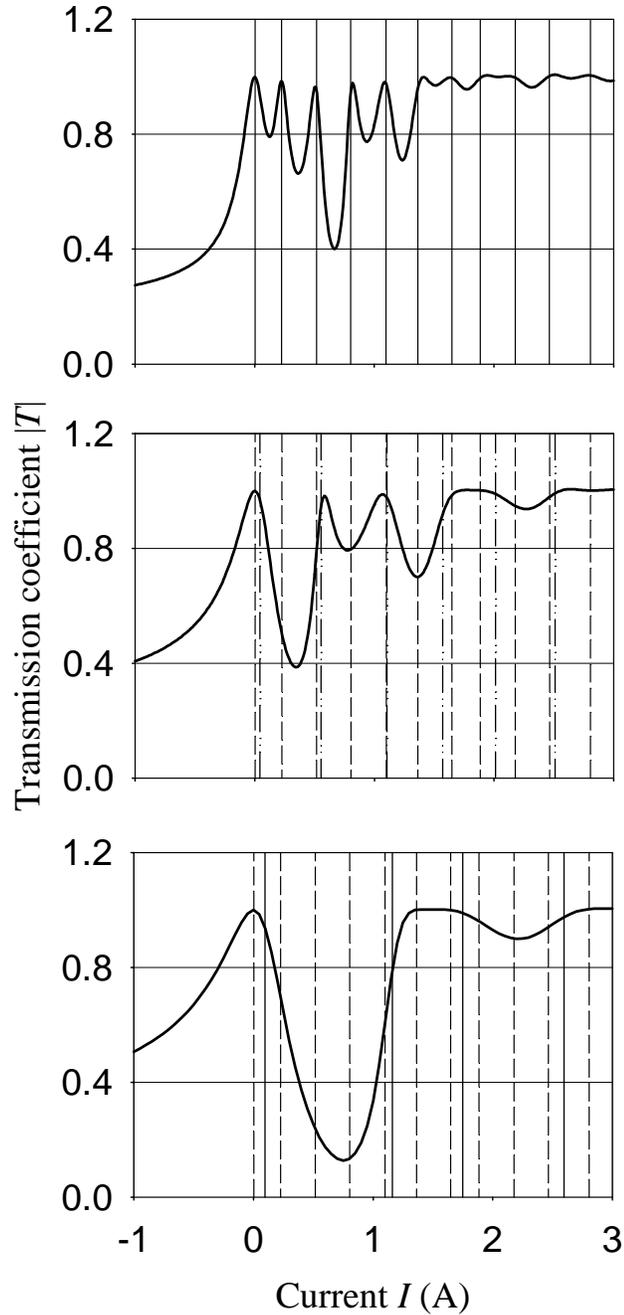

Fig. 7. Transmission coefficient vs. d.c. current magnitude for the inhomogeniety profiles (21). Lower panel: experimental inhomogeneity profile (4) ($t=50$ μm, $l_{eff}=25$ μm), middle panel: $t=100$ μm, $l_{eff}=25$ μm, upper panel: $t=200$ μm, $l_{eff}=25$ μm. Thin lines show positions of roots of (13). Dashed line: for the largest inhomogeneity $t=200$ μm, $l_{eff}=25$ μm, dash-dotted $t=200$ μm, $l_{eff}=25$ μm, solid line: $t=50$ μm, $l_{eff}=25$ μm. For convenience of comparison the dashed line is shown in all the panels.

The calculation using the experimental inhomogeneity profile of Eq. (4) is shown in the lower panel of Fig. 7. The positions of the roots of (13) nearly coincide with each fourth root for $t=200$ μm, and each second root for $t=100$ μm. No sharp resonance is visible and only plateaus of perfect transmission are formed. As seen from the upper panel of Fig. 4 the inverse eigenvalues are closely spaced for these currents. The values $\text{Im}(1/\lambda_n)$ determine the width of resonance lines, and the plateaus are formed by overlapping of neighboring resonances because of large resonance linewidths.

The quality factor of the resonances is $\text{Re}(1/\lambda_n)/\text{Im}(1/\lambda_n)$ and decreases with the length of the inhomogeneity. A linewidth of a loaded resonance is proportional to the coefficient of coupling of the resonator to the external waveguide and inversely proportional to the energy stored in it. In our case the coupling depends on the reflection coefficient of the inhomogeneity boundaries, whereas the stored energy is proportional to the inhomogeneity length. Since the reflection coefficient remains constant for all inhomogeneity profiles, the linewidths of resonances increases with decrease of $t$. For the same reason the resonances at smaller $I$ have smaller quality factors, since the boundary is more penetrable. In this case the resonances are loaded more efficiently by waves leaving the inhomogeneity.

Thus we have shown that the non-monotonic behavior of $|T(I)|$ in the regime of scattering is due to resonant scattering. The maximum of transmission at non zero $I$ is formed by overlapping of neighboring resonances because of large resonance linewidths and the distribution of the resonances as a function of $I$. The former effect is connected with the small length of the resonator.

Transmission and incident wavenumber

We now examine how the amplitude of the transmission coefficient in the minimum of transmission seen in Fig. 2 increases with the value of incident wavenumber $k_0$. The calculations shown in Fig. 8 indicate that the quality factors of resonances as well as the differences between the neighboring resonances $|\text{Re}(1/\lambda_{n+1}) - \text{Re}(1/\lambda_n)|$ decrease with increasing $k_0$. This gives a larger overlapping of neighbouring resonance lines, and hence a smaller variation of transmission between maxima and minima. The reason for the increase of resonance linewidths is that there is a smaller relative change of the wavenumber $|(k_i - k_0)/k_0|$ for the same magnitude of inhomogeneity when $k_0$ becomes larger. (Here $k_i$ is the wavenumber inside the inhomogeneity (Fig. 1c).) A smaller magnitude of $|(k_i - k_0)/k_0|$ means a smaller reflection from the inhomogeneity edges and the wave is less trapped by the inhomogeneity, corresponding to a larger resonance linewidth $|\text{Im}(1/\lambda_n)|$ and hence less structure in the transmission resonances.

This phenomenon also has an analogy to quantum mechanics. In [33] it is shown that for the square potential the amplitude of the transmitted wave in the minima of transmission between two consecutive resonances is proportional to $|E/V| = |-(\hbar k_0)^2/(2mV)|$. Thus it grows with the wavenumber of the incident wave.

With the increase of $|E/V|$ for a given potential $V$ (in our case with the increase of the quantity $|(1+\upsilon_0)/I\delta\nu(z)|$ for given $I$) the particle's energy is farther above the top of the potential well $E - V = 0$ (in our case the quantity $1 + \upsilon_0$ is farther from its value for the upper edge of BVMSW range inside the inhomogeneity $-I\delta\nu(z)$). The quantity

$|(E-V)/E|$ (BVMSW: $|[1+\upsilon_0 + I\delta v(z)]/(1+\upsilon_0)|$) becomes closer to 1, therefore $|(k_i - k_0)/k_0|$ becomes smaller. Therefore the reflection from the well edges decreases, resulting in less pronounced minima of transmission.

Thus with increase of $1+\upsilon_0$ the initial wavenumber $k_0$ grows. Therefore as the magnetostatic wave approaches the inhomogeneity, it is "energetically" farther from the prohibited zone $1+\upsilon_0 < 0$ and is less trapped as well as reflected back by the well.

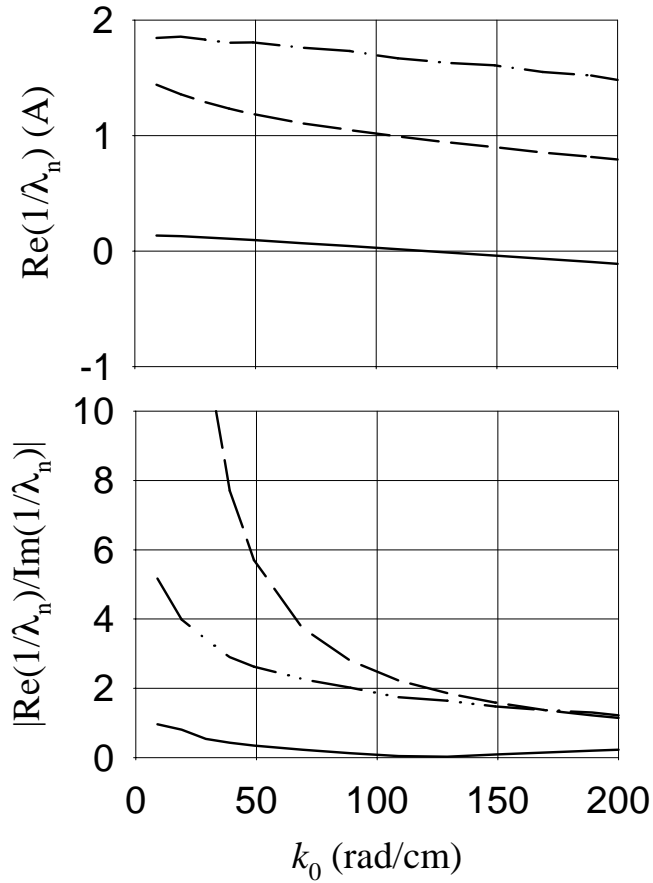

Fig. 8. Real parts of inverse eigenvalues (upper panel) and the "quality factors" for eigenvalues (lower panel) vs. the incident wavenumber . Solid line: the trivial eigenvalue n=0, dashed line: n=1, and dash-dotted line n=2.

Accumulated phase shift

Finally we discuss the magnitude of the additional accumulated phase shift as a function of the applied current. As seen from Figs. 2 and 3 for positive current values the experimental phase shift has a general tendency to decrease linearly with the current applied. We can obtain a simple formula for the phase shift in this region using the WKB approximation.

We define the local wavenumber $k(z)$ as a solution of the local dispersion relation $\chi(z,\omega)^{-1} - W(|k(z)|) = 0$, introduced above. With the same approximations used in Eqs. (33) and (47) we arrive at the expression for the local wavenumber. We find

$$k(z) = -\frac{2}{L}\left[1 + \upsilon_0(\omega) + I\delta\upsilon(z,\omega)\right] = -|k_0| - \frac{2}{L}I\delta\upsilon(z,\omega) \qquad . \tag{22}$$

Note that we accounted for the negative dispersion of BVMSW by putting a negative sign in front of the brackets. The additional accumulated phase is $\Delta\phi = \int_{-w/2}^{w/2} [k(z) - k_0]dz$. As a result, with the upper expression from Eq. (7) we arrive at an expression for $\Delta\phi$ in terms of $I$ and the profile shape:

$$\Delta\phi = \frac{4\pi}{5L}\eta(\omega)I \qquad . \tag{23}$$

The dash-dot-dotted line in the lower panel of Fig. 3b shows the phase of the transmission coefficient calculated using Eq. (23). The linear dependence of the phase accumulated on the height of the inhomogeneity is well described by the WKB approximation. The additional phase shift accumulated is primarily connected to the modified wavenumber of the transmitted wave in the barrier.

We note the numerical calculations shown in Fig. 3 are most nonlinear at small wavenumber. The greatest nonlinearity occurs at current magnitudes corresponding to maxima and minima of the transmission coefficient.

Our experiment showed that for the negative current direction the phase shift behaviour is very different. At small negative values of current the phase shift remains linear in $I$, but is nonlinear at larger currents. This can now be understood by noting that at small negative $I$ the incident wave is scattered from the inhomogeneity and phase accumulation is due to a modified wavenumber while inside the inhomogeneity. Because of the negative BVMSW dispersion, a decrease of the wavenumber by the inhomogeneity causes the additional phase shift to be positive.

At larger negative current values a zone prohibited for BVMSW propagation is formed. The zone length $l_p$ is the root of the equation $\upsilon_0(\omega) + \delta\upsilon(l_p/2, \omega)I = -1$. The zone length grows with the strength of applied current following the law $l_p(I) = \sqrt{\left[(r+d)^2 - \exp[-\eta(\omega)I/(5L)](r+d+L)^2\right] / \left[\exp[-\eta(\omega)I/(5L)]-1\right]}$. A plot of this is shown in the inset to Fig. 3b. Since the inhomogeneity is smooth, in front of and behind the zone there exist regions of reduced static field where the wave can propagate. In these two regions of allowed propagation the wave accumulates a phase shift. Tunnelling through the prohibited zone results in a negligible accumulated phase. A larger prohibited zone length does not therefore lead to a significant increase in phase shift. Hence the dependence on current is not linear in this large current limit.

We now prove this idea with a short calculation. We use the following procedure to calculate the additional accumulated phase shift in presence of the prohibited zone. As stated above, while deriving Eq. (22) from Eqs. (33) and (47), we assumed $k$ in Eq. (47) to be negative. The wavenumber $k(z)$ in Eq. (22) becomes positive for values of $z$ which define boundaries for the prohibited zone. Assuming that the accumulated phase in the prohibited zone is zero, we estimate the additional accumulated phase shift by subtracting an integral over the range of positive $k(z)$ values from the integral Eq. (23). This gives

$$\Delta\phi = \frac{2\eta(\omega)}{5L}\left[2\pi - \frac{1}{L}\int_{-l_p(I)/2}^{l_p(I)/2} Y(z)dz\right]I - k_0 l_p(I) \qquad . \tag{24}$$

The function $Y(z)$ is positive everywhere. Hence with increase of the current the dependence $\Delta\phi(I)$, Eq. (24), deviates from a straight line towards smaller $|\Delta\phi|$ values, which is in a qualitative agreement with Fig. 2. We calculate the magnitude of the current $I_t$, for which the prohibited zone begins to form. In the incident wavenumber range 50-200 rad/cm, $I_t$ varies from -0.088 A to -0.47 A. These values are in agreement with the change of curve character from linear to nonlinear in the lower panel of Fig. 2.

We also made calculations of the phase shift by using Eq. (24). As one sees from Fig. 3, the results are in a good agreement with the rigorous solution of the integral equation.

Thus in this section we have shown that the experimentally observed linear behaviour of the accumulated phase on the current applied for $I > 0$ is primarily connected to the local variation of wavenumber in the potential well. The observed tendency of saturation of $\Delta\phi(I)$ for large $I<0$ is connected with zero phase accumulation in the prohibited zone and the growth of the length of the prohibited zone with $I$.

## Conclusion

We have studied experimentally and theoretically the transmission of a dipole-dominated spin wave in a ferromagnetic film through a localised inhomogeneity in form of magnetic field produced by a dc current through a wire placed on the film surface. We show that the amplitude and phase can be simultaneously affected by the current induced field, a feature that will be relevant for logic based on spin wave transport.

The direction of the current creates either a barrier or well for spin wave transmission. We experimentally found that the current dependence of the amplitude of spin wave transmitted through the well inhomogeneity is non-monotonic. The dependence has a minimum and an additional maximum. The theory clarifies the origin of the maximum. It shows that the transmission of spin waves through the inhomogeneity can be considered as a scattering process and that the additional maximum is a scattering resonance.

A linear decrease of the phase of the transmitted wave on the height of the inhomogeneity was found experimentally in the regime of wave scattering from the field hump (well inhomogeneity). The theory and the experiment showed that the additional phase accumulation is primarily connected to the variation of spin wave wavenumber in the potential well. The non-monotonic resonance behaviour of amplitude and strong reflection from the barrier in the minimum of transmission do not result in a significant change of the overall character of the dependence. The phase dependence in the regime of wave scattering or tunnelling through the field dip was found to deviate from linear behaviour at a critical current. The critical current corresponds to the formation of a prohibited zone for spin wave propagation.

**Acknowledgment**

This work was supported in part by Deutsche Forschungsgemeinschaft (Project Hi 380/13), and by Australian Research Council Linkage Project LX0560336.

## Appendix A

To derive (1) we used a procedure as follows. First we assume a harmonic oscillatory motion for the magnetization and the dynamic magnetic field

$$\mathbf{m}(\mathbf{r},\omega) = \mathbf{m}(\mathbf{r})\exp(i\omega t), \quad \mathbf{h}(\mathbf{r},\omega) = \mathbf{h}(\mathbf{r})\exp(i\omega t) \quad . \tag{25}$$

Also we use a 1D-Green's function presentation for the dipole field of precessing magnetization in the film

$$\mathbf{h}_d(z) = \mathbf{G}(z,z') \otimes \mathbf{m}(z) , \tag{26}$$

where $\otimes$ denotes the convolution operation and $\mathbf{h}_d(z)$ and $\mathbf{m}(z)$ are the dynamic dipole field and the dynamic magnetization averaged through the film thickness [30]. All the components of the tensorial Green's function $\mathbf{G}$ can be found in [30]. For linear BVMSWs only the diagonal component $G_{xx}$ is important, since it is the only non-vanishing component, which induces a dipole field in the plane perpendicular to the direction of the film equilibrium magnetization. (Note, that the in-plane component of the dynamic magnetization $m_y$, being parallel to BVMSW wavefronts, produces no dipole field, if the film width tends to infinity compared to its thickness).

Thus

$$h_{dx}(z) = 4\pi G_{xx}(z,z') \otimes m_x(z) . \tag{27}$$

From the other hand the solution of the linearized Landau-Lifschitz equation has a form [28, 29]:

$$\mathbf{m}(z) = \hat{\chi}(\omega, z)\mathbf{h}_{eff}(z) \ . \tag{28}$$

The effective dynamic field $\mathbf{h}_{eff}$ in our case consists of the BVMSW dipole field and a microwave field of an external source exciting magnetization oscillations

$$\mathbf{h}_{eff}(z) = \mathbf{h}_d(z) + \mathbf{h}_s(z) \ . \tag{29}$$

(Other possible contributions to $\mathbf{h}_{eff}(z)$, if necessary, are taken into account in $\hat{\chi}(\omega, z)$.)

The excitation source can be of different nature, it might be a microstrip antenna with a microwave current or a dynamic field of any inhomogeneity in the film, e.g., a dipole field of another wave in a region of inhomogeneous static magnetic field. In both cases the external excitation field has only one component which affects the BVMSW dynamics. It is the *x*-component. Therefore

$$\mathbf{h}_{eff} = \mathbf{e}_x (h_{dx} + h_{sx}) \ , \tag{30}$$

where $\mathbf{e}_x$ is the unit vector in *x*-direction. With (30) equation (28) reduces to:

$$m_x(z) = \chi(\omega, z)\left[h_{dx}(z) + h_{sx}(z)\right] , \tag{31}$$

or, taking into account (27):

$$\chi(\omega, z)^{-1} m_x = 4\pi G_{xx}(z, z') \otimes m_x(z) + h_{sx}(z) \ . \tag{32}$$

Now we specify the form of the external field $h_{sx}(z)$. We assume it to be the microwave magnetic field of a line source of infinitesimally small width in the direction $z$ and situated at $z_0$. The amplitude of the microwave magnetic field is $A$. Under these assumptions Eq. (32) turns into (1).

**Appendix B**

Here we derive the Green's function of excitation of dipole-dominated spin waves in a ferromagnetic film by an external source, which enters (1).

First we make use of the fact that $\delta H_z(z) << H_s$ and expand the inverse of the diagonal component $\chi$ of the microwave susceptibility tensor:

$$\frac{\chi(z,\omega)^{-1}}{4\pi} = \upsilon_0(\omega) + \delta\upsilon(z,\omega)I + O\left[\left(\frac{I}{H_s}\right)^2\right], \tag{33}$$

where

$$v_0(\omega) = \frac{\chi(z = \pm\infty, \omega)^{-1}}{4\pi} = \frac{\omega_H^2 - \omega^2}{\omega_H \omega_M}, \tag{34}$$

and

$$\delta v(z, \omega) = \eta(\omega) Y(z). \tag{35}$$

Here

$$\eta(\omega) = \frac{1}{H_s}\left(\frac{2\omega_H}{\omega_M} - v_0(\omega)\right), \tag{36}$$

and $\omega_H = \gamma H_s$, $\omega_M = \gamma 4\pi M_S$, where $4\pi M_S = 1750$ Oe is the saturation magnetization of the YIG film used, and $\gamma = 2.82 \cdot 10^6$ Hz/Oe.

If $I$ is negative, $H(z)$ is reduced near the wire and $|\chi(z,\omega)^{-1}/(4\pi)|$ is increased in this region. A zone prohibited for BVMSW propagation exists were $|\chi(z,\omega)^{-1}/(4\pi)| > 1$. Tunneling can occur through this region, but not propagation [22]. If $|\chi(z,\omega)^{-1}|/(4\pi) < 1$ throughout the inhomogeneity region, an incident spin wave can scatter. This is realized for small $I>0$ such that BVMSW propagation is allowed in the inhomogeneity region.

Using Eqs. (3)-(8) the integral equation of motion Eq. (1) becomes

$$-\upsilon_0(\omega)m(z) + \int_{-\infty}^{+\infty} G_{xx}(z,z')m(z')dz' = I\delta\upsilon(z,\omega)m(z) + A'\delta(z-z_0) \quad, \tag{37}$$

where $A'$ is a new constant specifying the amplitude of the incident wave.

The source terms on the right hand side of Eq. (37) are independent. The solution of Eq. (37) is

$$m(z) = I\int_{-w/2}^{w/2} G_{exc}(z,z')\delta\upsilon(z',\omega)m(z')\mathrm{d}z' + A'G_{exc}(z,z_0), \tag{38}$$

where $G_{exc}(z,z')$ denotes the Green's function of excitation of dynamic magnetization by a point source located at $z'$.

If follows from (37) that in the BVMSW case it is the solution of equation, as follows:

$$\left[-\upsilon_0(\omega)\delta(z-z') + G_{xx}(z,z')\right] \otimes G_{exc}(z',z'') = \delta(z-z'') \quad. \tag{39}$$

The homogeneous equation (cp. (37))

$$\upsilon_0(\omega)m^0(z) - \int_{-\infty}^{+\infty} G_{xx'}(z,z')m^0(z')dz' = 0 \tag{40}$$

describes the BVMSW propagation in a homogeneously magnetized film. The equation represents an eigenvalue problem for the integral operator, in which $\upsilon_0(\omega)$ plays the role of the operator's eigenvalue. One easily finds that the operator has a continuous set of eigenfunctions in form of traveling waves

$$m^0(z) = \exp(ikz) \tag{41}$$

with arbitrary real wave numbers $k$. Substitution of (41) into the operator results in the expression for its eigenvalues $W(k)$:

$$W(k) = \frac{1}{|k|L}\left[\exp(-|k|L) - 1\right]. \tag{42}$$

As seen from the expression (42), the set of eigenvalues is doubly degenerate, to each eigenvalue $W$ correspond two eigenfunctions: $m^0(z) = \exp(i|k|z)$ and $m^0(z) = \exp(-i|k|z)$. The eigenfunctions represent two plane waves with frequency $\omega$ and wavenumbers $|k|$ and $-|k|$, traveling in opposite directions. Substitution of (41) and (42) into (40) results in the dispersion relation for BVMSW

$$\upsilon_0(\omega) - W(|k|) = 0 \quad. \tag{43}$$

As the plane waves represent the eigensolutions of the homogeneous equation (40), we search for the solution of the inhomogeneous equation (39) in form of a set of plane waves:

$$G_{exc}(z,z') = \int_{-\infty}^{\infty} g_k \exp[-ik(z-z')] dk \ . \tag{44}$$

Then, taking into account the result (42), we obtain

$$G_{exc}(z,z') = \frac{1}{2\pi} \int_{-\infty}^{\infty} \frac{\exp[-ik(z-z')]}{W(|k|) - \upsilon_0(\omega)} dk \ . \tag{45}$$

If $|k_0|$ is the wave vector value which satisfies the dispersion relation (43) for a given value of $\omega$, then we can rewrite (45), as follows

$$G_{exc}(z,z') = \frac{1}{2\pi} \int_{-\infty}^{\infty} \frac{\exp[-ik(z-z')]}{W(|k|) - W(|k_0|) - i\upsilon_0^{"}(\omega)} dk \ . \tag{46}$$

Here we phenomenologically introduced the magnetic losses by adding an imaginary part $\upsilon_0^{"}(\omega)$ to $\upsilon_0(\omega) = \chi_0(z,\omega)^{-1}/(4\pi)$.

Note that $\upsilon_0^{"}$ is positive. This follows from the expression for the microwave magnetic susceptibility tensor in the presence of magnetic losses [29]: $\chi = \chi' - i\chi"$, and $\chi" > 0$. Therefore $\chi^{-1} = (\chi' + i\chi")/(\chi'^2 + \chi"^2) \equiv 4\pi(\upsilon' + i\upsilon")$, and $\upsilon" > 0$.

The following condition is usually satisfied in experiments: $|k|L \ll 1$. Under this condition

$$W(|k|) \approx \frac{|k|L}{2} - 1. \tag{47}$$

This result allows one to obtain (46) in closed form

$$G_{exc}(z,z') = \frac{1}{\pi L}\Big[ 2\pi i \exp\left(ik_0^c |z-z'|\right) \\ + \exp\left(ik_0^c |z-z'|\right) E_1\left(ik_0^c |z-z'|\right) + \exp\left(-ik_0^c |z-z'|\right) E_1\left(-ik_0^c |z-z'|\right) \Big]. \tag{48}$$

Here

$$k_0^c = |k_0| + i\frac{2\upsilon_0^{''}}{L} \tag{49}$$

is the complex wavenumber of the wave excited by the source in a resonant way, and $E_1(z)$ is the exponential integral. It has a series presentation as

$$E_1(z) = -C - \ln(z) - \sum_{n=1}^{\infty} \frac{(-1)^n z^n}{nn!} \text{ [33]}.$$

The first term in the brackets of (48) represents traveling waves propagating in both directions from the point source. The second term, singular along the line *z=z'*, represents the near (reactive) field of the source. The reactive field is localized at the source and exhibits no retardation. Far away from the excitation source the reactive field vanishes and $G_{exc}(z,z')$ reduces to

$$G_{exc}(z,z') \cong \frac{2i}{L}\exp\left(ik_0^c |z-z'|\right), \quad |z-z'| \gg L. \tag{50}$$

Then last term of Eq. (38) reduces to $B\exp(ik_0^c z)$, where $B$ is a constant. $B$ represents a normalized amplitude of the excitation source, and we can set $B$ equal to 1 with no loss of generality. As a result the equation of motion of magnetization (38) takes its final form:

$$m(z) = I \int_{-w/2}^{w/2} G_{exc}(z,z') \delta \upsilon(z',\omega) m(z') \mathrm{d}z' + \exp(ik_0^c z). \tag{51}$$

An expression similar to the expression (50) was first obtained in a different way in [35]. Note that the whole wave factor in the far zone is $\exp(ik_0^c |z-z'|)\exp(i\omega t)$ (cp. (25)). Hence the wave exited at $z = -\infty$ and incident onto the inhomogeneity from the left side has the negative wave number $-|k_0|$, since it is $\exp\left[i\left(|k_0| + i2\upsilon_0''/L\right)z\right]$, rather than $\exp\left[-i\left(|k_0| + i2\upsilon_0''/L\right)z\right]$, which vanishes at $z = +\infty$. This unusual feature reflects the fact that BVMSW is a backward wave. Its phase velocity is anti-collinear to its group velocity. The direction of wave propagation is the direction of its group velocity, which in a passive medium coincides with the direction of decay of its amplitude. Thus in the direction of the group velocity, i.e. in the direction of incidence the phase accumulated by BVSMW on a propagation path $l$ $-|k_0|l$ is negative.

**Appendix C**

Here we derive an explicit transmission formula for small $I$. It follows from the discussion above that in the vicinity of $I = 0$ we may neglect contributions of the higher resonant term of the series Eq. (11) and keep only the zero order $\lambda_0$. We then have

$$T \simeq \frac{\int_{-w/2}^{w/2} \exp(ik_0^c z)\phi_0(z)\mathrm{d}z}{1 - I\lambda_0} u_0(w/2) \ . \qquad (52)$$

Reference to Figure 5 tells us that the inhomogeneity profile of Eq. (35) is a good approximation for the modulus of $\phi_0(z)$ and we set $\phi_0(z) = \delta v(z,\omega)\exp(ik_0|z|)$ for the phase. We can also approximate $u_0(w/2) \simeq N^2 \exp(ik_0 w)$, where $N$ is a constant determined by the normalization condition Eq. (12). Because $|u_0(z)|$ is close to a constant, we use

$$N^2 \simeq \int_{-w/2}^{w/2} \delta v(z,\omega)\exp(i2k_0|z|)\mathrm{d}z \qquad (53)$$

as an estimation of the norm. Then again, approximating $u_0(z)$ by the constant function with the phase modulation $\exp(ik_0|z|)$, and using the bi-orthogonality condition Eq. (12) we obtain:

$$\lambda_0 = \frac{1}{N^2} \int_{-w/2}^{w/2} dz \delta \upsilon(z,\omega) \exp(ik_0|z|) \int_{-w/2}^{w/2} \delta \upsilon(z',\omega) G_{exc}(z,z') \exp(ik_0|z'|) dz' \quad . \qquad (54)$$

The dash-dot-dotted lines in Fig. 6 show $T$ given by Eq. (52) along with the exact numerical results obtained using the eigenfunction expansion method. The $T(I)$ behavior at small $I$ (both positive and negative) is well described by Eq. (52). A discrepancy appears for large $I$ such that $|I| \gg |\text{Re}(1/\lambda_1) - \text{Re}(1/\lambda_0)|$ because the tails of higher transmission resonance lines become comparable with that of the lowest one. This indicates a transition to the tunneling regime, since the simultaneous out-of-resonance excitation of all the eigenmodes describes forced motion of magnetisation within the inhomogeneity, excited by a source at the inhomogeneity boundary.

.